\documentclass[twocolumn,aps,prb,showpacs,preprintnumbers,amsmath,amssymb]{revtex4}
\usepackage{float}
\usepackage{graphicx}  
\usepackage{array}
\usepackage{multirow}
\usepackage{graphicx}
\usepackage{dcolumn}   
\usepackage{bm}        
\usepackage{amssymb}   
\usepackage{amsmath}
\usepackage[dvipsnames,usenames]{color}
\usepackage{color}
\usepackage{comment}
\usepackage{gensymb}
\usepackage{natbib}
\begin{document}

\newcommand{\etal}{{\sl et al.}}
\newcommand{\ie}{{\sl i.e.}}
\newcommand{\sto}{SrTiO$_3$} 
\newcommand{\ef}{$E_{\rm F}$}
\newcommand{\eg}{$e_{g}$}
\newcommand{\tg}{$t_{2g}$}

\title{Role of the exchange-correlation functional on the structural, electronic and optical properties of cubic and
tetragonal SrTiO$_3$ including many-body effects}

\author{Vijaya Begum}
\email[]{vijaya.begum@uni-due.de}
\author{Markus Ernst Gruner}
\email[]{Markus.Gruner@uni-due.de}
\author{Rossitza Pentcheva}
\email[]{Rossitza.Pentcheva@uni-due.de}
\affiliation{Department of Physics and Center for Nanointegration Duisburg-Essen (CENIDE), University of Duisburg-Essen, Lotharstr. 1, 47057 Duisburg, Germany}
\date{\today}

\begin{abstract}
SrTiO$_3$ is a model perovskite compound with unique properties and technological relevance. At 105 K it undergoes a transition from a cubic to a tetragonal phase with  characteristic antiferrodistortive rotations of the TiO$_6$ octahedra. Here we study systematically the effect of  different exchange correlation functionals  on the structural, electronic and optical  properties of cubic and tetragonal STO by comparing the recently implemented strongly constrained and appropriately normed (SCAN) meta-GGA functional with the  generalized gradient approximation (PBE96 and PBEsol) and the hybrid functional (HSE06). SCAN is found to significantly improve the description of the structural properties, in particular  the rotational angle of the tetragonal phase, comparable to HSE06 at a computational cost similar to GGA. The addition of a Hubbard $U$-term (SCAN+$U$, $U=7.45$ eV) allows to achieve the  experimental band gap of 3.25 eV with a moderate increase in the lattice constant, whereas within GGA+$U$ the gap is underestimated even for high $U$ values. The effect of the exchange-correlation functional on the optical properties is progressively reduced from 1.5 eV variance in the onset of the spectrum in the independent particle picture to 0.3 eV upon  inclusion of many-body effects within the framework of the $GW$ approximation (single-shot $G_0W_0$) and  excitonic corrections by solving the Bethe-Salpeter equation (BSE).  Moreover, a model BSE approach is shown to reproduce the main  features of the optical spectrum at a lower cost compared to $G_0W_0$+BSE. Strong excitonic effects are found in agreement with previous results and their origin is analyzed based on the contributing interband transitions. Last but not least, the effect of the tetragonal distortion on the optical spectrum is discussed and compared to available experimental data.
\end{abstract}

\maketitle

\section{Introduction}

\begin{figure}[!htp]
\includegraphics[width=0.47\textwidth]{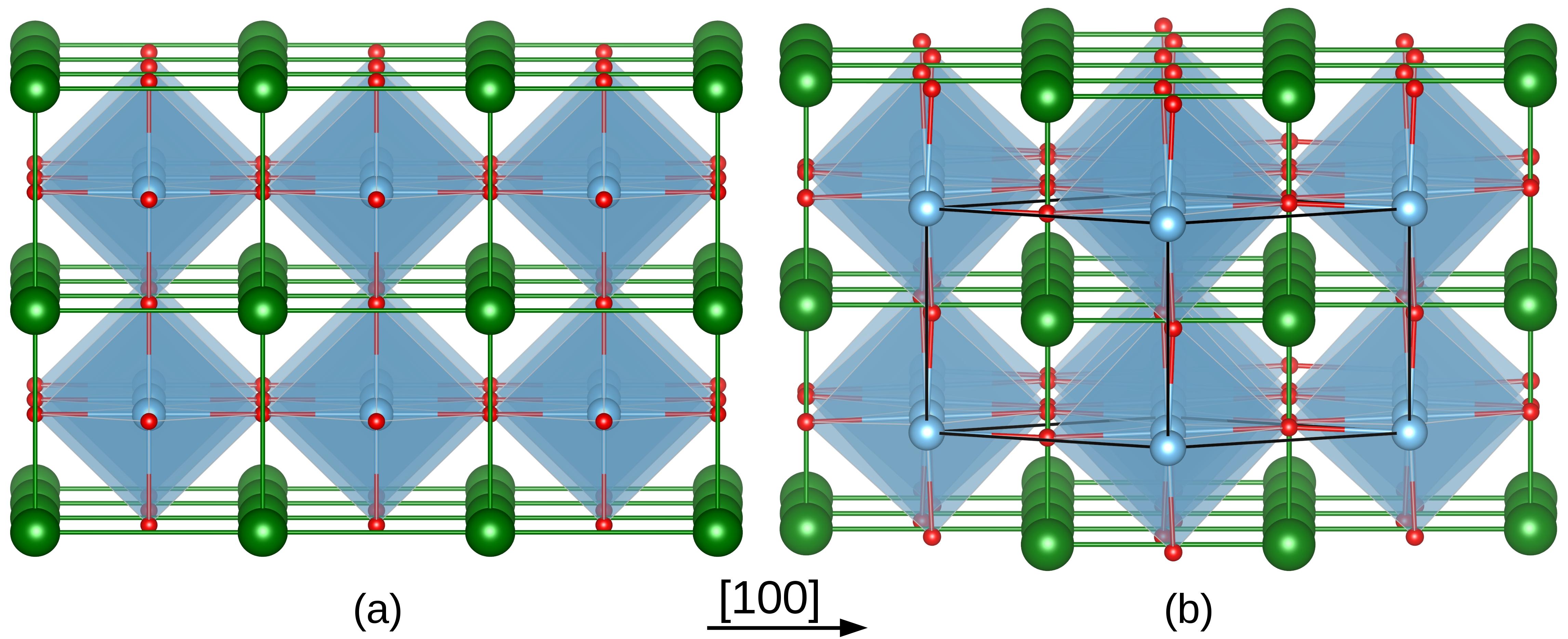}%
\caption{Top view of the cubic (a) and tetragonal (b) phase of SrTiO$_3$ with the antiferrodistortive rotation of the TiO$_6$ octahedra and the 10-atom simulation cell in the latter. Sr, Ti and O ions are shown in green, light blue and red, respectively. }
\label{Fig:struct}
\end{figure}

SrTiO$_3$ (STO) is a paradigmatic perovskite material. It is a quantum paraelectric where ferroelectricity (FE) is suppressed by quantum oscillations\cite{QPE}. A cubic to tetragonal transition takes place at 105 K, associated with a antiferrodistortive (AFD) rotation of the TiO$_6$ octahedra \cite{Fleury,Shirane,Unoki}, cf. Fig.  \ref{Fig:struct}b. Under strain it exhibits a competition between FE and AFD instabilities\cite{Aschauer,Vanderbilt}. Moreover, electric field or small doping leads to superconducting behavior\cite{Ueno}, whose origin is strongly debated. STO serves also as substrate material for high $T_{\rm C}$ superconductors and in oxide electronics\cite{Mannhart:10}. Despite being a band insulator in the bulk with a charge transfer type of a band gap between occupied O $2p$ states and empty Ti $3d$ states, STO  hosts a two-dimensional electron gas at its interface e.g. with LaAlO$_3$ \cite{Hwang} and at the surface\cite{Santander-Syro, Meevasana}. 

As a model perovskite compound, bulk STO has been intensively studied with different first principles methods\cite{Piskunov,Wahl,Heifets,Mellouhi}. The unique behavior and the strong interplay between structural distortions and electronic properties require a high accuracy in the theoretical description. Available local and semi-local functionals either underestimate \cite{Wahl,Mellouhi} (local density approximation (LDA)) or overestimate \cite{Wahl,Mellouhi} (generalized gradient approximation (GGA)) the lattice parameters and both underestimate strongly the band gap. Moreover, GGA in the implementation of Perdew, Burke and Enzerhof (PBE96)\cite{PBE} significantly overestimates the rotational angle of the AFD phase \cite{Wahl,Mellouhi}. On the other hand, hybrid functionals can improve the description of both the structural\cite{Piskunov,Wahl,Heifets,Mellouhi} and electronic \cite{Wahl,Heifets,Mellouhi}properties, but at a significantly higher computational cost. We take the opportunity to explore here the performance of the newly implemented meta-GGA SCAN exchange-correlation functional\cite{SCAN} and compare to the results obtained with  GGA (PBE96, PBEsol\cite{PBEsol}) and the screened hybrid functional of Heyd, Scuseria and Ernzerhof\cite{HSE06_1,HSE06_2}(HSE06). The SCAN functional has been shown to improve the description for a wide variety of materials such as van der Waals-bonded compounds as well as some oxides.\cite{SCAN2}

Besides the structural and electronic properties, the calculation of optical spectra allows a direct comparison with experimental data. Various experimental methods have been employed to obtain the optical spectrum of STO such as reflectivity\cite{Cardona}, ellipsometry\cite{Benthem}, x-ray absorption spectroscopy\cite{Lee}, among others. Here we will refer to the optical spectrum measured by Benthem, Els\"asser and French \cite{Benthem} for the cubic phase, using valence electron-energy loss and ultraviolet spectroscopy (VEELS and VUV) and to the  temperature-dependent ellipsometry data by Gogoi and Schmidt \cite{Gogoi} for the AFD phase.
\par
In early theoretical work the optical spectrum has been mostly calculated within the ``independent-particle'' (I.P.) picture \cite{Wang,Samantaray}. Although these calculations were able to reproduce the experiment to some extent, the I.P. picture fails to describe optical excitation correctly. The calculation of quasiparticle energies requires approaches beyond DFT, such as many-body perturbation theory (MBPT). Within the $GW$ approximation, introduced by Hedin\cite{Hedin}, the self-energy $\Sigma$ is computed as a product of the single particle Green's function $G$ and the screened Coulomb interaction $W$, $\Sigma=iGW$. Several studies have performed previously $GW$ calculations (mostly single shot $G_{0}W_{0}$) on STO \cite{Ergonenc,Cappellini,Kim}. While these studies are based on input from  LDA,
here we study systematically the effect of the exchange-correlation functional on the optical spectrum, using GGA (PBE96), meta-GGA (SCAN) and a hybrid functional (HSE06). We note that previous studies have addressed the role of the starting point for the $G_0W_0$ calculation mainly on the band gaps by comparing LDA vs. LDA+$U$\cite{Jiang} for rare earth sesquioxides or hybrid functionals for elemental and binary semiconductors\cite{Fuchs07} and SrPdO$_3$\cite{He2014}. Beyond the $GW$ approximation, electron-hole interactions may significantly alter the spectrum. These can be accounted for by solving the Bethe-Salpether equation\cite{BSE_1,BSE_2} (BSE). Two previous $G_{0}W_{0}+$BSE studies  on cubic STO \cite{Sponza, Gogoi2} show significant improvements of the spectrum, pointing towards the important role of excitonic contributions for this material, whereas some deviations still remain. In this work, we analyze the contribution of interband transitions to the excitonic features and also compare the performance of a model BSE approach to describe the spectrum. Moreover, we explore the effect of the tetragonal distortion on the optical spectrum that to our knowledge has not been addressed so far from first principles. 
\par
The paper is structured as follows: In Section \ref{Sec:Compdetail} we discuss the computational details. Section \ref{sec:strel} deals with the influence of the exchange-correlation functional on the structural and electronic properties of cubic and tetragonal STO, also including the effect of a Hubbard $U$ term within SCAN+$U$ and PBE96+$U$. In Section \ref{Sec:OptProp} we discuss the optical spectrum of cubic STO calculated using different exchange-correlations functionals (GGA-PBE96, SCAN and HSE06) and different levels of description: I.P. picture, $G_{0}W_{0}$ and BSE. The results are compared to the experimental and previous theoretical studies. Insight into the origin of the peaks in the optical spectrum is gained by analyzing the interband transitions with the eigenvectors obtained as solution of the BSE equation. Moreover, the validity of a model-BSE approach is verified. We furthermore explore the influence of the tetragonal distortion on the optical spectrum i.e. the imaginary part of the dielectric tensor  and compare to the experimental results of Gogoi and Schmidt \cite{Gogoi}.

\section{Computational details} \label{Sec:Compdetail}
The DFT calculations are performed with the \textit{Vienna Ab initio Simulation Package} (VASP)\cite{VASP1,VASP2}, using the projector augmented wave (PAW) method \cite{Pseudopotential}. As mentioned above, for the  exchange-correlation functional we have used the generalized gradient approximation in the implementation of Perdew Burke and Ernzerhof \cite{PBE}, PBEsol \cite{PBEsol1,PBEsol2}, the SCAN meta-GGA \cite{SCAN} and the HSE06 hybrid functional\cite{HSE06_1,HSE06_2}. Beyond GGA, in which the exchange-correlation energy density depends on the electron density and its gradient, the SCAN functional includes the positive orbital kinetic energy densities. It is the only semi-local exchange-correlation functional which satisfies the 17 exact constraints and is appropriately normed, i.e. it accurately calculates interactions in rare-gas atoms and unbonded system, but also shows a very promising performance for a variety of materials including van der Waals-bonded systems as well as some oxides\cite{SCAN2}. Due to the underestimation of the band gap within GGA and meta-GGA, we have additionally explored the effect of an on-site Coulomb repulsion parameter\cite{Liechtenstein}, where we have applied the implementation of Dudarev\cite{Dudarev} with an effective $U_{\rm eff}=U-J$ with $U$ and $J$ being the Coulomb repulsion and exchange terms. In hybrid functionals a fraction of exact nonlocal exchange is included, which is separated into a long-range (lr) and a short-range (sr) part in real space:  
\begin{equation} \label{eq:HSE06}
E_\mathrm{XC}^\mathrm{HSE}= \alpha E_\mathrm{x}^ \mathrm{sr,\mu} + (1-\alpha)E_\mathrm{x}^\mathrm{PBE,sr,\mu} + E_\mathrm{x}^\mathrm{PBE,lr,\mu} + E_\mathrm{c}^\mathrm{PBE}.
\end{equation}
\par
The range separation parameter $\mu$ controls the characteristic distance at which the long-range nonlocal interaction becomes negligible. For HSE06\cite{Krukau}, $\mu$=0.11 a.u.$^{-1}$=0.207$\, \textrm{\AA}^{-1}$ and the mixing parameter is $\alpha=$25\%. 

Cubic/tetragonal STO was modeled in a 5-/10-atom unit cell with a total of 40/80 valence electrons (Ti: $3s^2 3p^6 4s^2 3d^2$, Sr: $4s^2 4p^6 5s^2$ and O: $2s^2 2p^4$). We used the GW PAW potentials recommended for excited state properties. A  $\Gamma$-centered  11$\times$11$\times$11/7$\times$7$\times$7 $\mathbf{k}$-mesh was used for the 5-/10-atom cell, unless otherwise specified. For comparison to the tetragonal phase, in some cases the cubic phase was also calculated using a 10-atom unit cell. The cut-off energy for the plane waves is 650 eV. For the AFD structure the initial data was taken from Ref.\cite{Jauch} measured at 50 K with lattice parameters $a^* =b^*=5.507$\AA\ $=\sqrt{2}a$ and $c^*=7.796$\AA\ and a  rotational angle of $2.1 \degree$. To reduce the computational demand, instead of this 20-atom cell we have used a reduced primitive triclinic 10-atom cell shown in Fig.  \ref{Fig:struct} with $a=b=c=5.510$\AA, $\alpha=\beta=120.034^{\circ}$ and $\gamma=89.942^{\circ}$, marked in Fig.  \ref{Fig:struct}b.
\par
For the single-shot $G_{0}W_{0}$ and BSE calculation, 100 frequency grid points were selected, with a cut-off energy of 650 eV and a total number of 192/384 bands for the 5/10 atom cell. To consider excitonic effects the BSE was solved within the Tamm-Dancoff approximation \cite{TDA} on top of the $G_{0}W_{0}$ quasiparticle calculation. In the BSE calculation the number of valence/conduction bands were 9/11 for the five atom cell and 18/22 for the ten atom cell. The optical spectrum is plotted for the $\varepsilon_{xx}$ of the imaginary part of the dielectric function $\varepsilon_{2}$, unless otherwise mentioned. A Gaussian broadening of 0.3 eV is used for the independent particle (I.P.) and $G_{0}W_{0}$ spectrum and 0.1 eV for the BSE spectrum.
\par

All the structures are visualized with VESTA\cite{Vesta}. For the band structure calculation the Wannier90\cite{Wannier90} package is used. The band structure path has been determined using AFLOW \cite{AFLOW}.
\par

\section{Results: Structural and electronic properties}
\label{sec:strel}
\subsection{Cubic phase}
\label{sec:cSTO}
\begin{table*}
\caption{\label{tab:Table1}Comparison of the lattice constant, bulk modulus and indirect ($R-\Gamma$) and direct ($\Gamma-\Gamma$) band gap calculated with different functionals for the cubic phase of STO.}
\begin{ruledtabular}
\begin{tabular}{ccccccccc}
\textrm{}&
\textrm{}&
\textrm{PBE96}&
\textrm{PBEsol}&
\textrm{SCAN}&
\textrm{HSE06}&
\textrm{B3LYP}&
\textrm{B3PW}&
\textrm{Experiment}\\
\colrule
$a_{0} \,$({\AA}) &Present & 3.938 & 3.896 & 3.901 & 3.896 & & & 3.905\footnotemark[2]\\[0.25ex]
 & Ref.\citenum{Wahl} & 3.943 & 3.898 & & 3.904 \\[0.25ex]
 & Ref.\citenum{Mellouhi} & 3.941 & 3.897 & & 3.902  \\[0.25ex]
 & Ref.\citenum{Piskunov} & 3.94 & & & & 3.94 & 3.90\\[0.25ex]
 & Ref.\citenum{Heifets} & & & & & & 3.912\\[0.25ex]
 \hline
$B_{0} \,$(GPa) & Present & 167 & 185 & 192 & 191 & & & 179\footnotemark[1]\\[0.25ex]
 & Ref.\citenum{Wahl} & 168 & 185 & & 192 & \\[0.25ex]
 & Ref.\citenum{Mellouhi} & 169 & 184 & & 193 & \\[0.25ex]
 & Ref.\citenum{Piskunov} & 169 & & & & 177 & 177 \\[0.25ex]
 \hline
$R-\Gamma$($\Gamma-\Gamma$) (eV) & Present & 1.81 (2.18) & 1.83 (2.19) & 2.26 (2.64) & 3.35 (3.73)& & & 3.25 (3.75)\footnotemark[3]\\[0.25ex]
 & Ref.\citenum{Wahl} & 1.80 (2.18) & 1.82 (2.18)& & 3.07 (3.47)\\[0.25ex]
 & Ref.\citenum{Mellouhi} & 1.74 (2.11) & 1.75 (2.10) & & 3.20 (3.59)\\[0.25ex]
 & Ref.\citenum{Piskunov} & 1.99 (2.35) & & & & 3.57 (3.89) & 3.63 (3.96)\\
\end{tabular}
\end{ruledtabular}
\footnotetext[1]{Reference \citenum{Hellwege}.}
\footnotetext[2]{Reference \citenum{Cao}.}
\footnotetext[3]{Reference \citenum{Benthem}.}
\end{table*}

At first we discuss the structural properties of cubic STO (space group \textit{Pm$\bar{3}$m}). The lattice constants and bulk moduli  obtained with different functionals are listed in Table \ref{tab:Table1}. With PBE96 the equilibrium lattice constant (3.938~\AA) is $\sim 1\%$ larger, while the values obtained with PBEsol, SCAN and HSE06 are very close to the experimental value of 3.905\cite{Cao}. The results obtained within PBE96, PBEsol and HSE06 agree with previous studies \cite{Wahl,Mellouhi,Piskunov,Heifets}. On the other hand the bulk modulus is underestimated within PBE96 (167 GPa) and overestimated by SCAN and HSE06 (192 and 191 GPa, respectively), while PBEsol gives the closest value to experiment (179 GPa)\cite{Hellwege}. 
Overall, SCAN predicts structural properties similar to the hybrid functional HSE06, but with a much lower computational effort, comparable to GGA (PBE96).

\begin{figure}[htp!]
\includegraphics[width=0.48\textwidth]{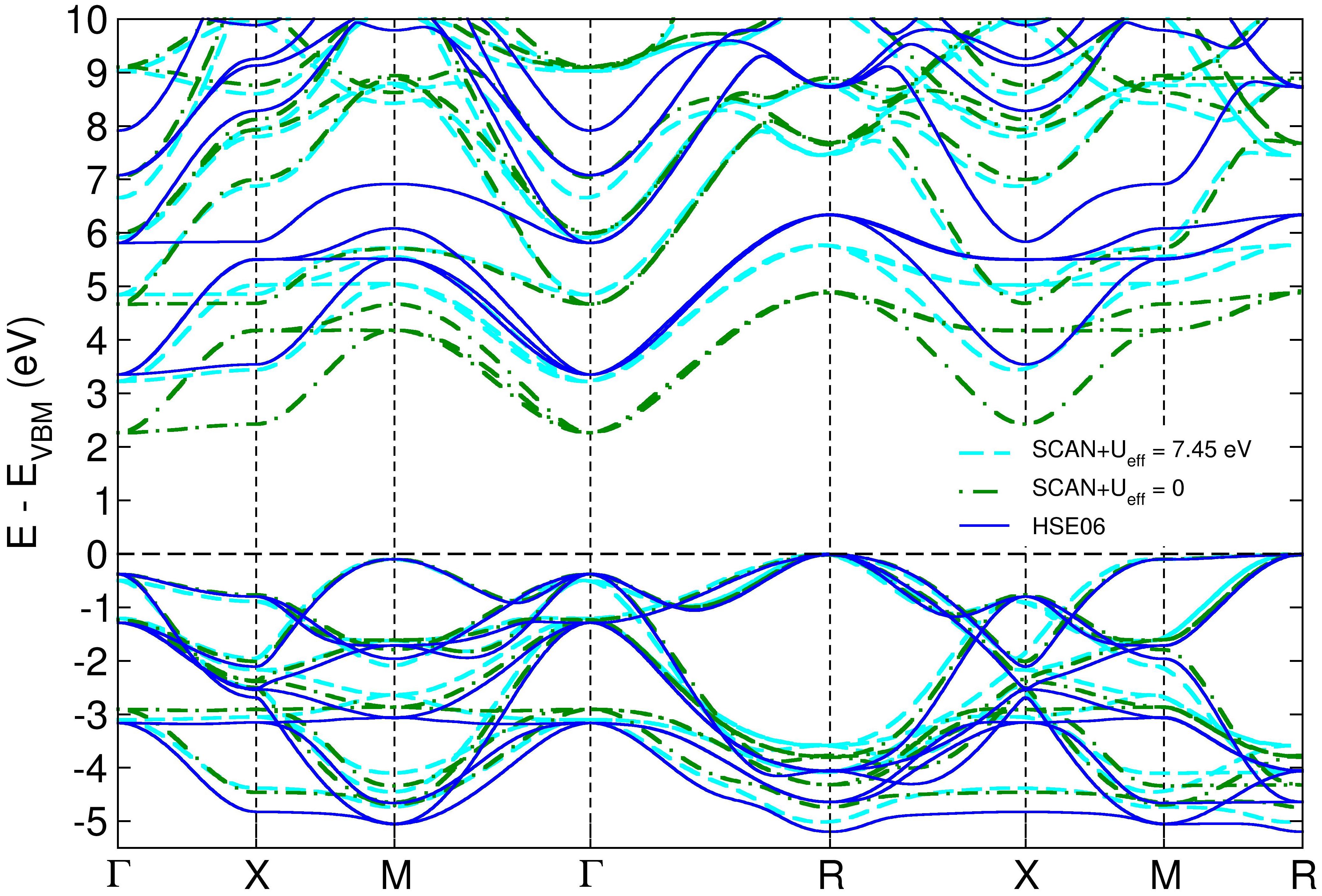}
\caption{Comparison of the band structure along the high symmetry directions with SCAN functional without and with $U_\text{eff}$ and HSE06. For SCAN with $U_\text{eff}=7.45$ eV the band gap is close to the HSE06 and the experimental value.}
\label{Fig:BandSCAN}
\end{figure}
\begin{figure}[htp!]
\includegraphics[width=0.48\textwidth]{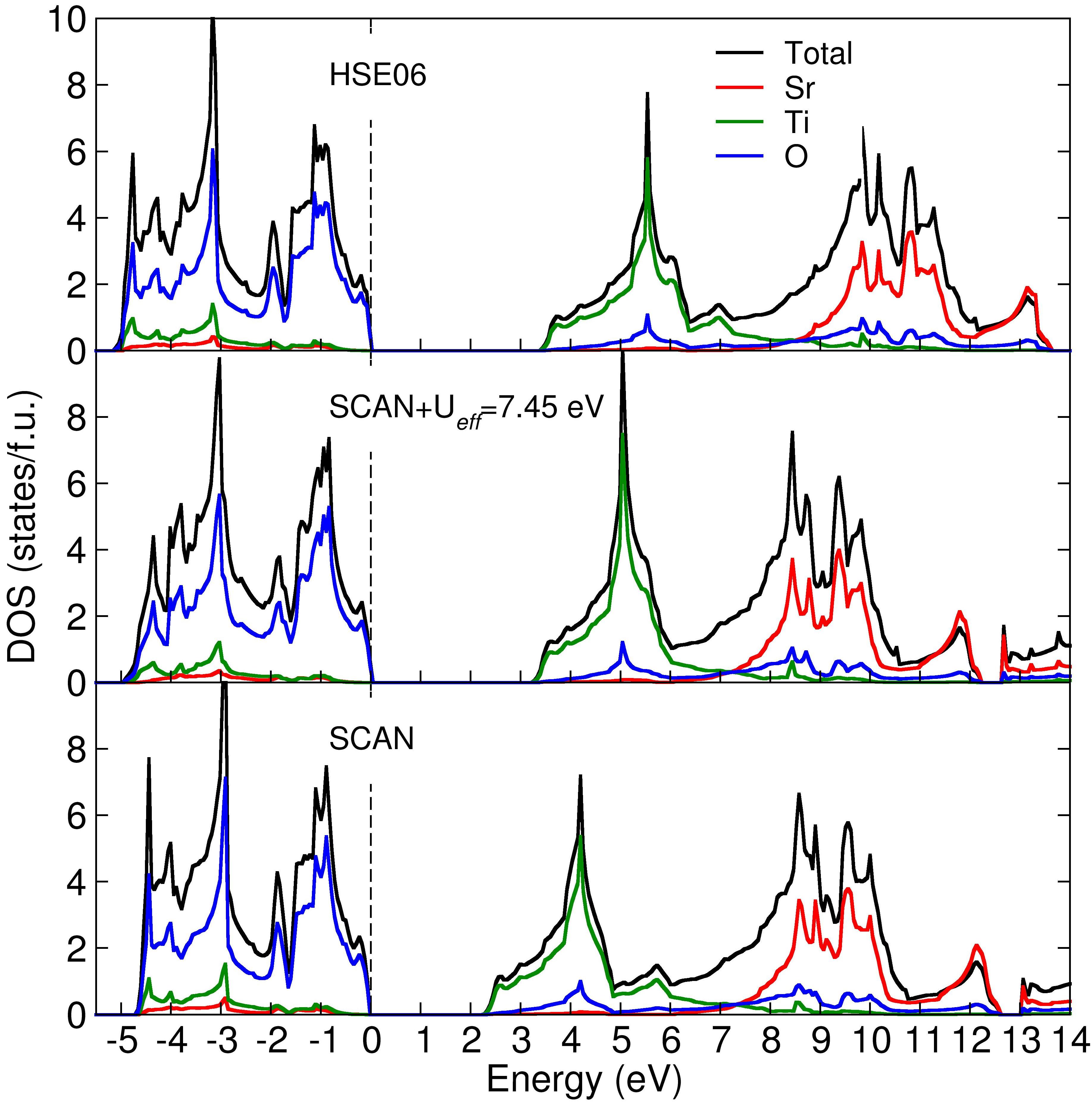}
\caption{Projected density of states (PDOS) obtained with HSE06 and SCAN without and with {$U_\text{eff}$}.}
\label{Fig:PDOS}
\end{figure} 
The  band structure obtained with SCAN and HSE06 is displayed in Fig. \ref{Fig:BandSCAN}. The valence band within SCAN ranges from -4.7 eV to 0.0 eV and  is somewhat broader for the HSE06 calculation. The valence band maximum (VBM) is at R, whereas the conduction band minimum (CBM) is at $\Gamma$. The values for the direct ($\Gamma$-$\Gamma$) and indirect (R - $\Gamma$) band gap obtained with the different functionals are given in Table \ref{tab:Table1}. With PBE96 the indirect and direct band gaps are 1.81 and 2.18 eV, respectively, $44\%$ and $42\%$ smaller than the experimental values. Similar values (1.83 and 2.19 eV) are obtained within PBEsol, consistent with Wahl et al. \cite{Wahl}. SCAN shows a notable improvement with indirect and direct band gaps of 2.26 and 2.64 eV, respectively, that reduces the underestimation to $30\%$.
For HSE06 the band gap values (3.35 and 3.73 eV) are slightly overestimated but are closest to the experimental values (3.25 and 3.75 eV) and in agreement with 
a previous study \cite{Mellouhi}. Other hybrid functionals such as B3LYP\cite{B3LYP} and B3PW\cite{B3PW} strongly overestimate the band gap \cite{Heifets}.
\par
Additionally, the projected  density of states (PDOS) in Fig. \ref{Fig:PDOS} reveals the contribution of each ion. In agreement with previous studies, the valence band is dominated by O $2p$ states, whereas the conduction band minimum comprises Ti $t_{2g}$ states (total width of $\sim 2.5$ eV), followed by a narrower Ti $e_{g}$ band, consistent with the octahedral coordination of Ti. Finally, a broader unoccupied Sr $4d$ and $5s$ band emerges between 7 and 12 eV. With HSE06 this band is shifted about 1 eV higher.  

\subsection{Effect of $U_\text{eff}$ on cubic STO}
\begin{figure}[htp!]
\includegraphics[width=0.5\textwidth]{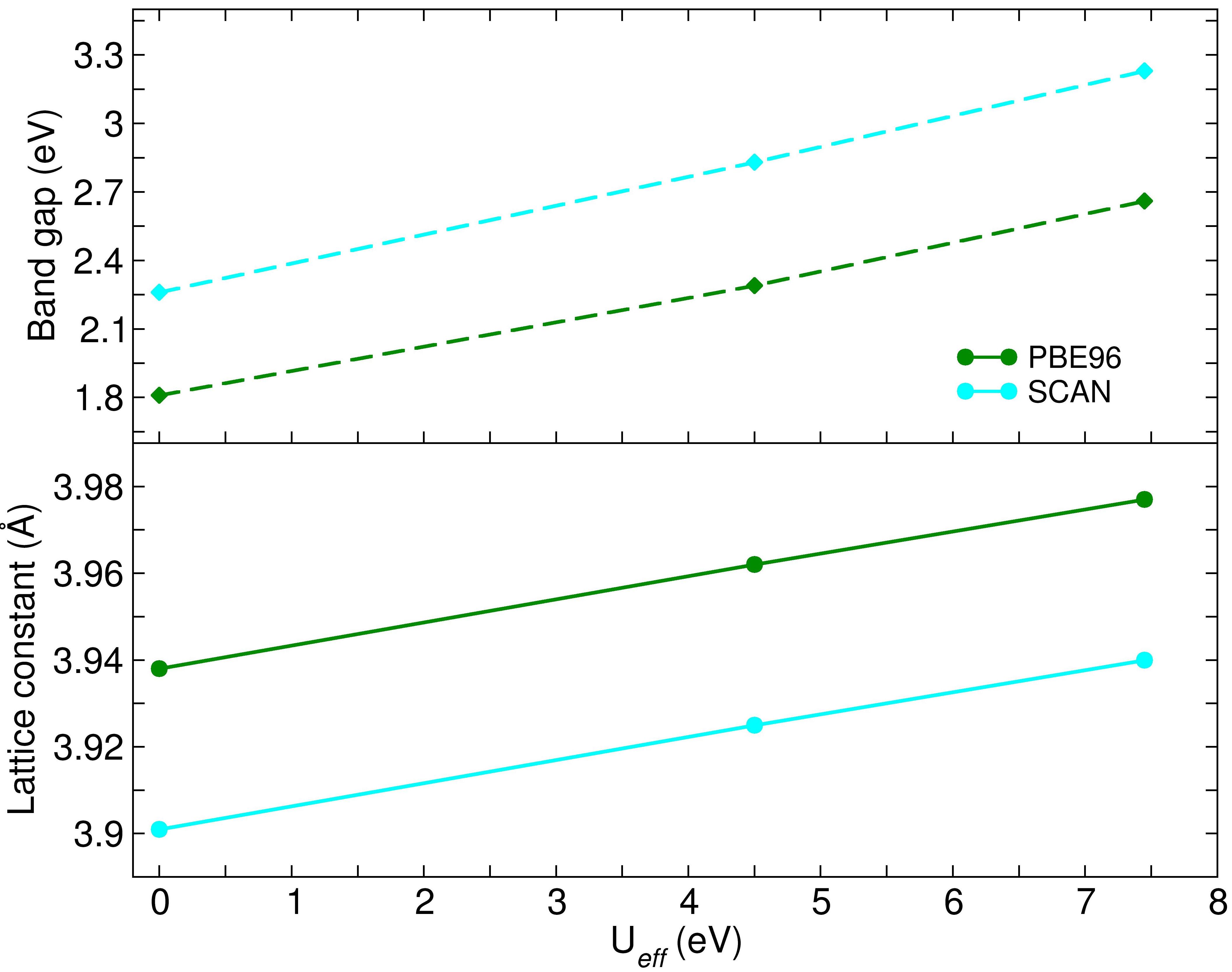}
\caption{Lattice constant and band gap of the cubic phase calculated with PBE96$+U$ and SCAN$+U$ as a function of $U_\text{eff}$. }
\label{fig:Ucbg}
\end{figure} 
\par
\begin{table}[h]
\caption{\label{tab:Ucbg} Comparison of the lattice constant and band gap for PBE96 and SCAN without and with $U_\text{eff}$.}
\begin{ruledtabular}
\begin{tabular}{lccccc}
\textrm{$U_\text{eff}$ in eV}&
\textrm{}&
\textrm{0}&
\textrm{4.5}&
\textrm{7.45}&
\textrm{Experiment}\\
\colrule
\multirow{2}{3em}{$a_{0}$ ({\AA})} & PBE96 & 3.938 & 3.962 & 3.977 & 3.890\footnotemark[1], 3.900\footnotemark[2]\\
 & SCAN & 3.901 & 3.925 & 3.940 &\\
\colrule
\multirow{2}{4em}{$E_g$ (eV) ($R-\Gamma$)} & PBE96 & 1.81 & 2.29 & 2.66 & 3.25\footnotemark[3]\\
 & SCAN & 2.26 & 2.83 & 3.23 &\\
\colrule
\multirow{2}{4em}{$E_g$ (eV) ($\Gamma-\Gamma$)} & PBE96 & 2.18 & 2.75 & 3.17 & 3.75\footnotemark[3]\\
 & SCAN & 2.64 & 3.28 & 3.72 &\\
\end{tabular}
\end{ruledtabular}
\footnotetext[1]{Reference \citenum{Hellwege}.}
\footnotetext[2]{Reference \citenum{Cao}.}
\footnotetext[3]{Reference \citenum{Benthem}.}
\end{table}
\begin{table*}
\caption{\label{tab:AFDprop}Comparison of the structural properties of the AFD phase with different functionals along with previous data. Here, a = $a^*/\sqrt{2}$ and c/a = $c^*/(\sqrt{2}a^*)$, where $a^*, c^*$ are the lattice parameters of the 20-atom cell. $\theta$ is the angle of the octahedral rotation, $\Delta E=E_\text{cubic}-E_\text{AFD}$ is the energy gain between cubic and tetragonal phase and $E_g$ gives the band gap, the values on brackets corresponds to the direct gap in the cubic phase. 
}
\begin{ruledtabular}
\begin{tabular}{ccccccccc} 
\textrm{}&
\textrm{}&
\textrm{PBE96}&
\textrm{PBEsol}&
\textrm{SCAN}&
\textrm{HSE06}&
\textrm{B3PW}&
\textrm{Experiment}\\
\colrule
\multirow{4}{3em}{$a$ \,({\AA})} & Present & 3.930 & 3.885 & 3.899 & 3.895 & & \multirow{4}{6em}{3.898\footnotemark[1] (65 K)}\\[0.25ex]
 & Ref.\citenum{Wahl} & 3.933 & 3.886 & & 3.900 & & & \\[0.25ex]
 & Ref.\citenum{Mellouhi} & 3.937 & 3.889 & & 3.90 & & \\[0.25ex]
 & Ref.\citenum{Heifets} & & & & & 3.910 & \\[0.25ex]
\hline
\multirow{4}{3em}{$c/a$} & Present & 1.004 & 1.006 & 1.001 & 1.001 & & \multirow{4}{7em}{1.0009\footnotemark[2] (10 K)}\\[0.25ex]
 & Ref.\citenum{Wahl} & 1.004 & 1.006 & & 1.001 & & \\[0.25ex]
 & Ref.\citenum{Mellouhi} & 1.0032 & 1.004 & & 1.0012 & & \\[0.25ex]
 & Ref.\citenum{Heifets} & & & & & 1.0006 & \\[0.25ex]
\hline
\multirow{4}{5em}{$\theta$ \,($\deg$)} & Present & 5.0 & 5.64 & 2.4 & 2.5 & & \multirow{4}{7em}{2.1\footnotemark[3] (4.2 K), 2.01\footnotemark[4] (50 K)}\\[0.25ex]
 & Ref.\citenum{Wahl} & 4.74 & 5.31 & & 2.63 & & \\[0.25ex]
 & Ref.\citenum{Mellouhi} & 3.54 & 3.81 & & 2.01 & & \\[0.25ex]
 & Ref.\citenum{Heifets} & & & & & 1.95& \\[0.25ex]
\hline
\multirow{4}{5em}{$\Delta E$ \,(meV/f.u.)}& Present & 4.5 & 6.6 & 0.8 & 0.34 & & \\[0.25ex]
 & Ref.\citenum{Wahl} & 3.5 & 5.5 & & 1.0 & & \\[0.25ex]
 & Ref.\citenum{Mellouhi} & 2.135 & 0.11 & & 0.0875 & & \\[0.25ex]
 & Ref.\citenum{Heifets} & & & & & 0.21 & \\[0.25ex]
\hline
\multirow{3}{7em}{$E_g$ (eV) \,$R-\Gamma$ ($\Gamma-\Gamma$)} & Present & 1.91 (2.21) & 1.95 (2.23) & 2.29 (2.65) & 3.38 (3.73) & & \multirow{3}{8em}{3.199 (indirect), 3.78 (direct)\footnotemark[5]}\\[0.25ex]
 & Ref.\citenum{Wahl}                     & 1.79 (2.12) & 1.93 (2.21) & & 3.11 (3.48) & & \\[0.25ex]
\end{tabular}
\end{ruledtabular}
\footnotetext[1]{Reference \citenum{Cao}.}
\footnotetext[2]{Reference \citenum{Unoki}.}
\footnotetext[3]{Reference \citenum{Heidemann}.}
\footnotetext[4]{Reference \citenum{Jauch}.}
\footnotetext[5]{Reference \citenum{Gogoi}, gaps measured at 4.2 K.}
\end{table*}

Compared to PBE96, SCAN improves both the structural properties and the band gap. Nevertheless, the latter is still significantly underestimated w.r.t. the experimental value. Therefore, we have  explored the influence of an on-site Hubbard term $U_\text{eff}$ applied on the empty Ti $3d$ states  on the structural and electronic properties. The results for PBE96$+U$ and SCAN$+U$ are summarized in Table \ref{tab:Ucbg} and Fig. \ref{fig:Ucbg}. The latter shows the well-known effect of a monotonic increase of both the lattice constant and the band gap with $U$. However, since the lattice constant with SCAN is very close to the experimental one for $U_\text{eff}=0$ eV, the overall overestimation is  only 1\% for  $U_\text{eff}=7.45$ eV, similar to the PBE96 value at $U_\text{eff}=0$ eV. In contrast, for PBE96$+U$, $U_\text{eff}=7.45$ eV, the lattice constant is 2$\%$ larger than the experimental one. Regarding the band gap the SCAN value is already higher (2.26 eV) than the PBE96 value (1.81 eV). SCAN$+U$ with $U_\text{eff}=7.45$ eV reaches nearly the experimental value, whereas the gap with PBE96$+U$ for the same $U_\text{eff}$  is still 0.6 eV smaller. The band structure and PDOS for SCAN$+U$ with $U_\text{eff}=7.45$ is shown in Fig. \ref{Fig:BandSCAN} and Fig. \ref{Fig:PDOS}. Interestingly, $U$ influences mainly the Ti $t_{2g}$ states, leading to an upward shift of the conduction band minimum, while the effect on the $e_{g}$ states is weaker. 
 
\par
Overall, within SCAN+$U$ the experimental band gap can be achieved. At the same time, the overestimation of the lattice constant is moderate compared to PBE96+$U$. 

\subsection{Structural and electronic properties of AFD phase}\label{sec:tSTO}


STO undergoes a phase transition from the cubic to a tetragonal phase\cite{Fleury,Shirane} at 105 K, as shown in Fig. \ref{Fig:struct}, which is characterized by rotations of neighbouring TiO$_6$ octahedra around the $c$-axis in antiphase to each other, described by ($a^0, a^0, c^-$) according to Glaser's notation. Unoki and Sakudo\cite{Unoki}  reported an octahedron rotation angle of $1.4 \degree$ at $77$ K and  $2.1 \degree$ at $4.2$ K. The results for the AFD phase obtained with the different functionals are summarized in Table \ref{tab:AFDprop}. The trends for the lattice parameter are similar to the cubic phase: while PBE96 overestimates the lattice constant by $\sim1\%$, PBEsol, SCAN and HSE06 give results very close to  experiment. 

PBE96 (1.004) and PBEsol (1.006) overestimate the $c/a$ ratio, in agreement with Wahl \textit{et al.}\cite{Wahl}. On the other hand, SCAN and HSE06 (1.001) lie very close to the experimental value and to the results with other hybrid functionals such as B3PW \cite{Heifets}. Most importantly, the rotational angle is strongly overestimated by PBE96 ($5.0 \degree$) and PBEsol ($5.64 \degree$), more than doubled compared to experiment. In contrast,   SCAN ($2.4 \degree$) and HSE06 ($2.5 \degree$)  render a significant improvement over PBE96 and PBEsol. As we will see below, this influences substantially the electronic properties.  

The degree of tetragonal distortion correlates also with the energy gain per formula unit (f.u.) between the two phases. Due to the strong overestimation of the rotation angle, with PBE96 (4.5 meV/f.u.) and PBEsol (6.6 meV/f.u.) the energy difference is significantly larger than for SCAN (0.8 meV/f.u.) and HSE06 (0.34 meV/f.u.). The PBE96 result is in agreement with previous studies \cite{Wahl,Mellouhi}.    

\begin{figure*}[ht!]
\includegraphics[width=0.9\textwidth]{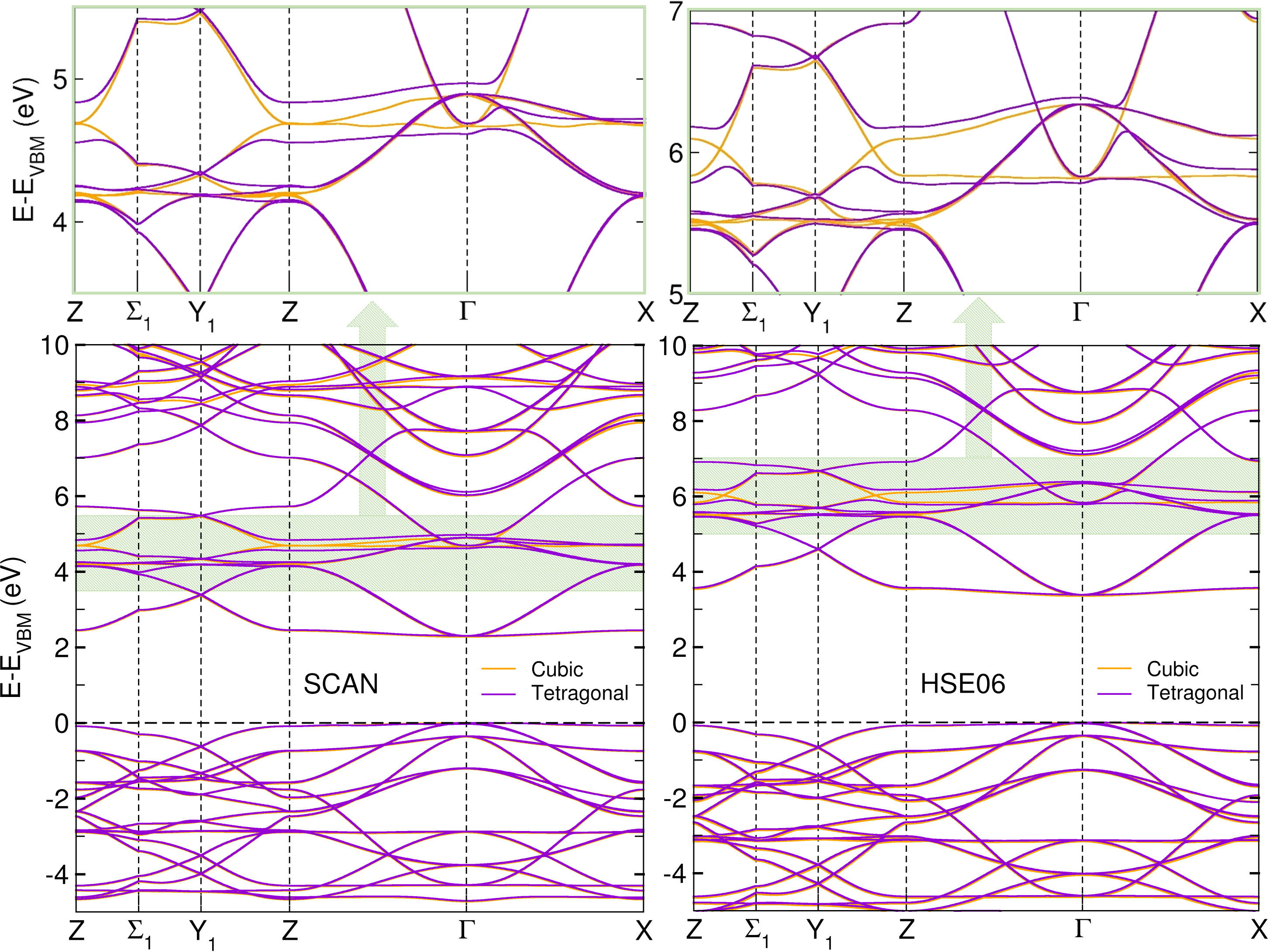}%
\caption{\label{Fig:BS-CubAFD}Comparison of the band structure of the AFD and cubic phase calculated with SCAN and HSE06 in the 10-atom unit cell. The top panels present  a magnified section of the green shaded area in the energy range 3.5 to 5.5 eV and 5.0 to 7.0 eV for SCAN and HSE06, respectively,  showing the splitting of the bands in the AFD phase predominantly along $\Gamma$-Z.}
\end{figure*}

As can be seen from Table \ref{tab:AFDprop}, the the AFD distortion leads to  a slight enhancement of the band gap by 84 and 115 meV for PBE96 and PBEsol. For SCAN and HSE06 the values are smaller, 22 and 25 meV, respectively, owing to the smaller rotational angle. The slight change is of similar value but opposite in sign to the measurement of Gogoi and Schmidt \cite{Gogoi}  at 4.2 K (3.199 eV). This may be  related to thermal expansion and vibrational effects that are not considered here. We note that the experimental values in Ref. \citenum{Gogoi} show a nonmonotonic dependence on temperature with increase up to 105 K and a subsequent decrease. We will address this later when we discuss the optical properties in Section \ref{Sec:OptProp}.

A comparison of the band structure of the cubic and AFD phases, calculated in the 10-atom u.c. is displayed in Fig. \ref{Fig:BS-CubAFD}. The larger unit cell results in a downfolding of the Brillouin zone, i.e. the former R-point now coincides with $\Gamma$, resulting in a \textit{direct} band gap for the AFD phase. Overall, the changes in the valence band and the lower conduction bands are minute, stronger changes are observed at $\sim 5$ eV above the Fermi level, where a pronounced splitting of bands occurs along $\Gamma$-Z, especially around Z. Somewhat weaker effects are observed above 7 eV, where the Sr $4d$ states are dominant. We note that the overestimation of the rotational angle leads to a much stronger band splitting within PBE96 (not shown here).

\section{RESULTS: Optical properties}\label{Sec:OptProp}
\subsection{Cubic phase}
\begin{figure}[!htp]
\includegraphics[scale=0.095]{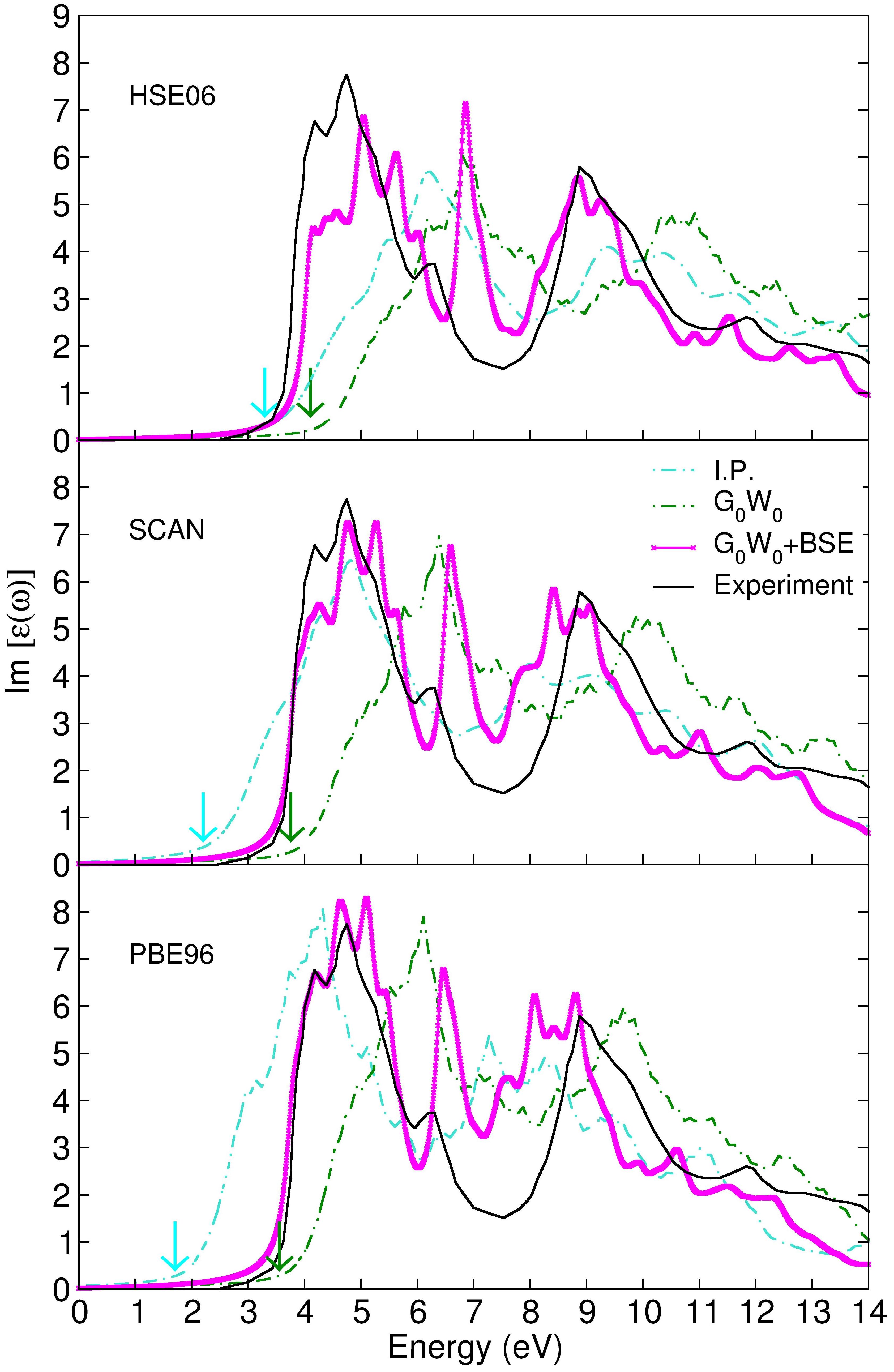}%
\caption{\label{Fig:OptSpe-Exc}Optical spectrum obtained within the I.P. picture, $G_0W_0$ and BSE correction for the cubic phase of STO, starting with different exchange-correlation functionals, PBE96, SCAN and HSE06. For comparison the experimental spectrum \cite{Benthem} is shown. The arrows indicate the indirect band gaps for the respective functional prior (cyan) and after (green) the $G_0W_0$ calculation.}
\end{figure}

In this Section we discuss the optical spectrum of cubic STO obtained using different exchange-correlation functionals (PBE96, SCAN and HSE06). We remind that the previous study of Sponza \textit{et al}.\cite{Sponza} used LDA as a starting point. The spectrum obtained within the I.P. picture is compared to the ones including many-body effects within the $G_0W_0$ approximation and excitonic effects by solving the Bethe-Salpeter equation. The calculated optical spectra are illustrated in Fig. \ref{Fig:OptSpe-Exc} together with the experimental results of Benthem, Els\"asser and French\cite{Benthem}. The experimental spectrum contains three main features: a first set of peaks with an onset slightly above the experimental band gap of 3.25 eV, extending up to 6 eV, a smaller peak at around 6.3 eV and another broad peak between 7.5 and 11 eV.

The spectrum obtained within I.P. with the different functionals reproduces the two main peaks. However, their positions depend strongly on the functional chosen: the onset of the spectrum is at $\sim 2.0$ eV (PBE96), $\sim 2.5$ eV (SCAN) and $\sim 3.5$ eV (HSE06), respectively, and correlates with  the size of the band gap with the respective functional (cf. Table \ref{tab:Table1}). The maximum of the first peak appears at 4.3, 4.8 and 6.5 eV, respectively. For SCAN it coincides with the first peak in the experimental spectrum. A similar trend is observed for the second peak which extends from 7.0 to 10 eV (PBE96), 7.5 to 11 eV (SCAN) and 8.5 to 12.5 eV (HSE06). The maximum of the experimental peak appears between the SCAN and HSE06 ones.

The main effect of $G_0W_0$ is a blueshift of the spectrum which is strongest for PBE96 ($\sim 2.0$ eV), and decreases for SCAN (1.5 eV) and HSE06 (1 eV). Although the differences in the three spectra are noticeably reduced, they are still shifted w.r.t. each other (however much less than in the I.P. case): the onset with PBE96 as a starting point is at 4.0 eV, followed by SCAN (4.3 eV) and HSE06 (4.8 eV). Thus the $G_0W_0$ spectrum with PBE96 as a starting point lies about 1 eV higher than the experimental one and the ones with SCAN$+G_0W_0$ and HSE06$+G_0W_0$ are even further shifted to higher energies. A comparison of the band gaps  before and after $G_0W_0$ in Table \ref{tab:ExcGWbg} shows a substantial overcorrection with $G_0W_0$ for all the three functionals under consideration: 3.64 eV (PBE96), 3.73 eV (SCAN) and 4.07 eV (HSE06). 
\begin{table}[!htp]
\caption{\label{tab:ExcGWbg} Comparison of the band gap (indirect/direct) of the cubic phase in the  I.P. and $G_0W_0$ approximation with different starting functionals.}
\begin{ruledtabular}
\begin{tabular}{lcccc}
\textrm{}&
\textrm{E$_{xc}$}&
\textrm{I.P.}&
\textrm{G$_{0}$W$_{0}$}&
\textrm{Experiment}\\
\colrule
\multirow{3}{5em}{$E_g$ (eV) $R-\Gamma$($\Gamma-\Gamma$)} & PBE96 & 1.81(2.18) & 3.64(4.00) & \multirow{3}{4em}{3.25(3.75)\footnotemark[1]}\\
 & SCAN & 2.26(2.64) & 3.73(4.09) &\\
 & HSE06 & 3.35(3.73) & 4.07(4.44) &\\
\end{tabular}
\end{ruledtabular}
\footnotetext[1]{Reference \citenum{Benthem}.}
\end{table}

\begin{figure*}[t!]
\includegraphics[scale=0.105]{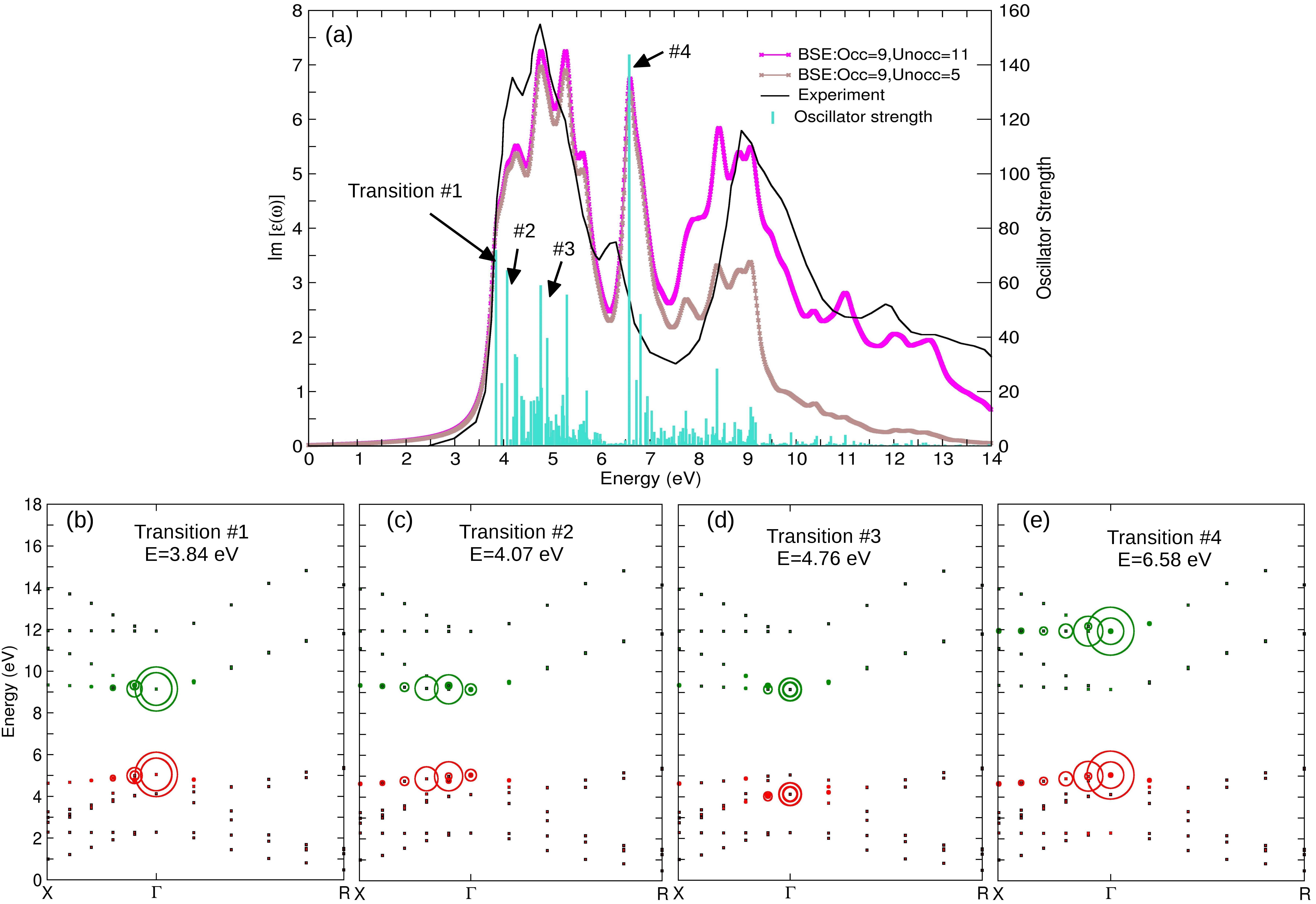}%
\caption{(a) Optical spectrum obtained with BSE with 11 and 5 unoccupied states, respectively. For the latter case the oscillator strength (with uniform scaling) is given. For comparison the experimental spectrum \cite{Benthem} is shown. (b-e) The electron-hole contribution of the marked transitions in reciprocal space where the radius of the circle corresponds to the magnitude of the electron-hole coupling coefficient.}
\label{Fig:SCANoscill}
\end{figure*}

The inclusion of the BSE correction has a substantial effect on the spectrum indicating strong excitonic contributions for STO, consistent with Sponza et al. \cite{Sponza}. Overall, BSE leads to a significant improvement of the agreement to the experimental spectrum and among the three exchange-correlation functionals. Due to the strong spectral weight transfer, the onsets of the three theoretical spectra are redshifted and lie in a much narrower range between 3.5 and 3.8 eV, the PBE96$+G_0W_0+$BSE starts somewhat earlier, while SCAN$+G_0W_0+$BSE is almost on top of the experimental onset and the HSE06$+G_0W_0+$BSE curve is at slightly higher energies. This relative shift remains also for the higher energy features. While PBE96 and SCAN show now best agreement to experiment w.r.t. the first peak, HSE06 renders best description of the third peak, indicating a superior description of the empty Sr $4d$ states compared to the other functionals. A prominent effect of BSE is the appearance of a sharp peak at 6.4 (PBE96), 6.6 (SCAN) and 7.0 eV (HSE06), that may be associated with the smaller peak in the experimental spectrum at 6.3 eV but is much more pronounced. Sponza et al. \cite{Sponza} have discussed the role of off-diagonal terms to the screening to reduce the peak. The height may also be influenced by dynamic screening (beyond $\omega=0$), electron-phonon or polaronic effects. 

\par
To get a deeper insight into the origin of peaks contributing to the optical spectrum, we have plotted in Fig. \ref{Fig:SCANoscill}a the respective oscillator strengths. While previous assignments of peaks were based on the band structure and a successive reduction of the number of bands contributing to the spectrum\cite{Sponza}, we  have   further analyzed the first four most prominent transitions with the aid of the BSE eigenvectors expressed in the electron-hole product basis\cite{mBSE} $\Phi^i=\sum_{c,v,\mathbf{k}}A^i_{c,v,\mathbf{k}}\phi_{c,\mathbf{k}}\phi_{v,\mathbf{k}}$,  starting from the SCAN calculation. In order to alleviate the computational cost and memory requirement for the BSE fatband calculation, we have used a reduced number of five unoccupied bands. The comparison with the spectrum with 11 unoccupied bands in Fig. \ref{Fig:SCANoscill} shows that this is sufficient to describe correctly the  relevant peaks up to 7 eV. In Fig. \ref{Fig:SCANoscill}(b-e), we plot  the coefficients  $|A^i_{c,v,\mathbf{k}} |$ of the exciton wave functions for the four transitions along high symmetry lines in reciprocal space, where the radius of the circle reflects the magnitude of a particular e-h pair contribution. In particular, the absolute magnitude of the unitless coupling coefficient can be extracted from the projection on the y-axis, multiplying the value by 500. 
The first transition (cf. Fig. \ref{Fig:SCANoscill}b) is localized at $\Gamma$ and corresponds to the direct gap of 3.85 eV between the top of the O $2p$ bands and the bottom of the conduction band comprised by Ti $t_{2g}$ states, leading to a shoulder in the optical spectrum, in agreement with the experimental analysis \cite{Gogoi}. The binding energy of this bound exciton is 0.246 eV in close agreement with 0.22 eV obtained previously\cite{Sponza,Gogoi2}. We note that by fitting the experimental data to Elliot's formula for a Wannier-Mott exciton, Gogoi and Schmidt \cite{Gogoi} obtained a weaker binding of 22 meV and a broadening of 40 meV. The second transition (Fig. \ref{Fig:SCANoscill}c) corresponds to the first peak at 4.07 eV and involves the same states, but is more dispersive along $\Gamma-X$. The third prominent transition (Fig. \ref{Fig:SCANoscill}d) at 4.76 eV is between the next (lower lying) set of O $2p$ states and the Ti $t_{2g}$ states at the conduction band minimum, again localized strongly at $\Gamma$. A transition at 5.20 eV exhibits only weak coupling along $\Gamma-M$ and is not shown here. The fourth transition (Fig. \ref{Fig:SCANoscill}e) we have considered is the excitonic peak at E$=$6.58 eV which involves the O $2p$ states at VBM and the Ti $e_{g}$ states, consistent with the analysis of Ref. \citenum{Sponza}. This transition is again more delocalized along the $\Gamma$-X direction and due to the low dispersion of the contributing O $2p$ and in particular Ti \eg\ states results in a high peak. We determine a binding energy of 0.185 eV for this exciton. Considering the similar octahedral environment of Ti in anatase or rutile, which implies similar $t_{2g}$ - $e_g$ splitting, this peak may be common to titanites, e.g. a similar peak has been observed at 6.12 eV in anatase TiO$_2$\cite{Yong2016}, but its exact origin needs to be investigated in future studies.

\texttt{\begin{figure*}[t!]
\centering
\includegraphics[scale=0.125]{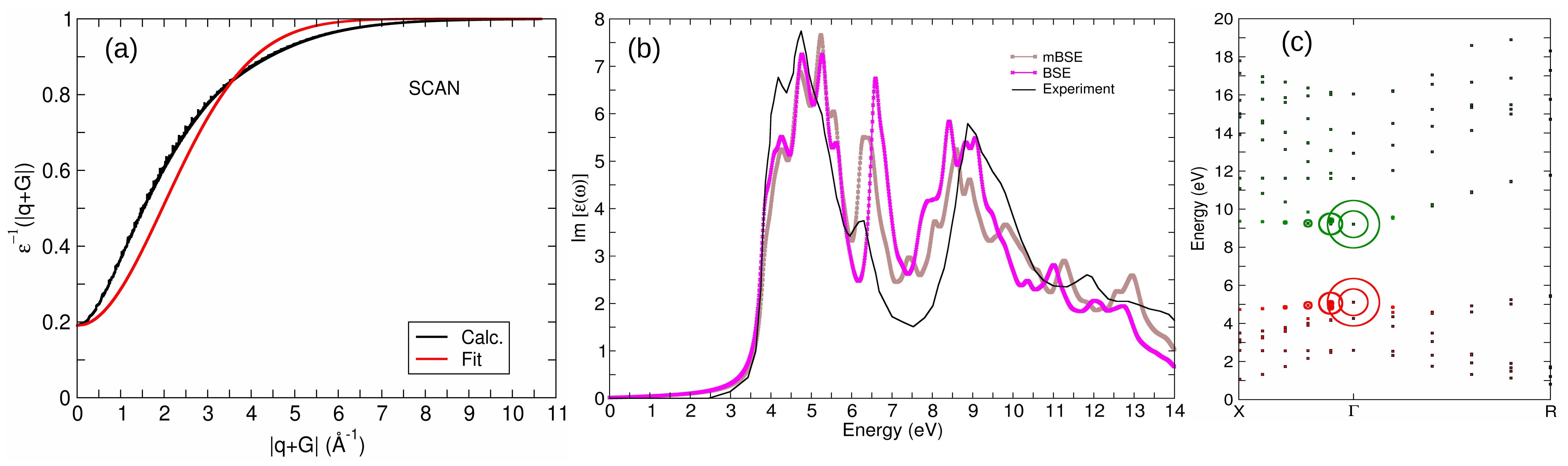}
\caption{(a) Inverse of the dielectric function $\epsilon^{-1}$ from the G$_{0}$W$_0$ calculation as a function of $\vert$\textbf{k+G}$\vert$ and the corresponding fit according to Eq. \ref{Eqn:mBSE}, (b) comparison of the $G_0W_0+$BSE and mBSE optical spectrum and (c) the first bright state from mBSE. The radius of the circle corresponds to the electron-hole coupling coefficient and have the same scale as Fig. \ref{Fig:SCANoscill}.}
\label{Fig:Excbrightst}
\end{figure*}}

\subsubsection{Model BSE}\label{Sec:mBSE}
Alternative to the complete $GW$+BSE calculation, we have also employed  a less computationally involved analytical model for the static screening \cite{mBSE,Bechstedt,Fuchs}. In this model BSE (mBSE) the imaginary part of the dielectric constant is described as:
\begin{equation}
\varepsilon^{-1}_{\mathbf{k+G}} = 1 - (1 - \varepsilon_{\infty}^{-1})e^{\dfrac{-(2\pi \vert \mathbf{k}+\mathbf{G} \vert)^{2}}{4\lambda^{2}}} .
\label{Eqn:mBSE}
\end{equation}  
where $\varepsilon_{\infty}$ is the ion-clamped static dielectric function, $\lambda$ is the range separation parameter obtained by fitting the curve to the screened Coulomb kernel diagonal values obtained from the $G_{0}W_{0}$ calculation and \textbf{G} is the lattice vector. Fig. \ref{Fig:Excbrightst}a shows the inverse of the dielectric function  from $G_0W_0$ and the fit to the model in Eq. \ref{Eqn:mBSE}. The parameters, obtained from the fit to the $G_{0}W_{0}$ dielectric function,  $\lambda=8.865$ \AA$^{-1}$ and $\varepsilon_{\infty}^{-1}=0.190$, are used as input for the mBSE calculation, which is carried out starting from the SCAN one-particle energies, applying the scissors operator with a shift corresponding to the difference between the  $G_{0}W_{0}$ and  SCAN band gaps. A $\Gamma$-centered 11$\times$11$\times$11 $\mathbf{k}$-mesh was used. The spectra from the full $G_0W_0$+BSE and the DFT+mBSE calculations are shown in Fig. \ref{Fig:Excbrightst}b. The features up to 6 eV are similar in energy and magnitude. In the mBSE calculation the excitonic peak at 6.5 eV is slightly shifted  to lower energy and  lower in magnitude, whereas the feature above 7.5 eV is shifted by $\sim 0.4$ eV to higher energy. Fig. \ref{Fig:Excbrightst}c shows the e-h pair contribution for the first exciton. In agreement with BSE [cf. Fig. \ref{Fig:SCANoscill}(b)], the transition is dominated by O $2p$ at VBM and Ti $t_{2g}$ bands at CBM and is localized at the $\Gamma$ point. The binding energy of the first exciton (0.233 eV) is close to the $G_0W_0+$BSE value (0.246 eV). Overall, the main characteristics of the optical spectrum for cubic STO are well reproduced in the mBSE approach. However, in other cases as e.g. for Ruddlesden-Popper iridates \cite{Liu} the agreement between the mBSE and full BSE calculation may be less satisfactory, due to the differences in position of the one-particle and quasiparticle energies.
\par

\begin{figure*}[t!]
\includegraphics[scale=0.12]{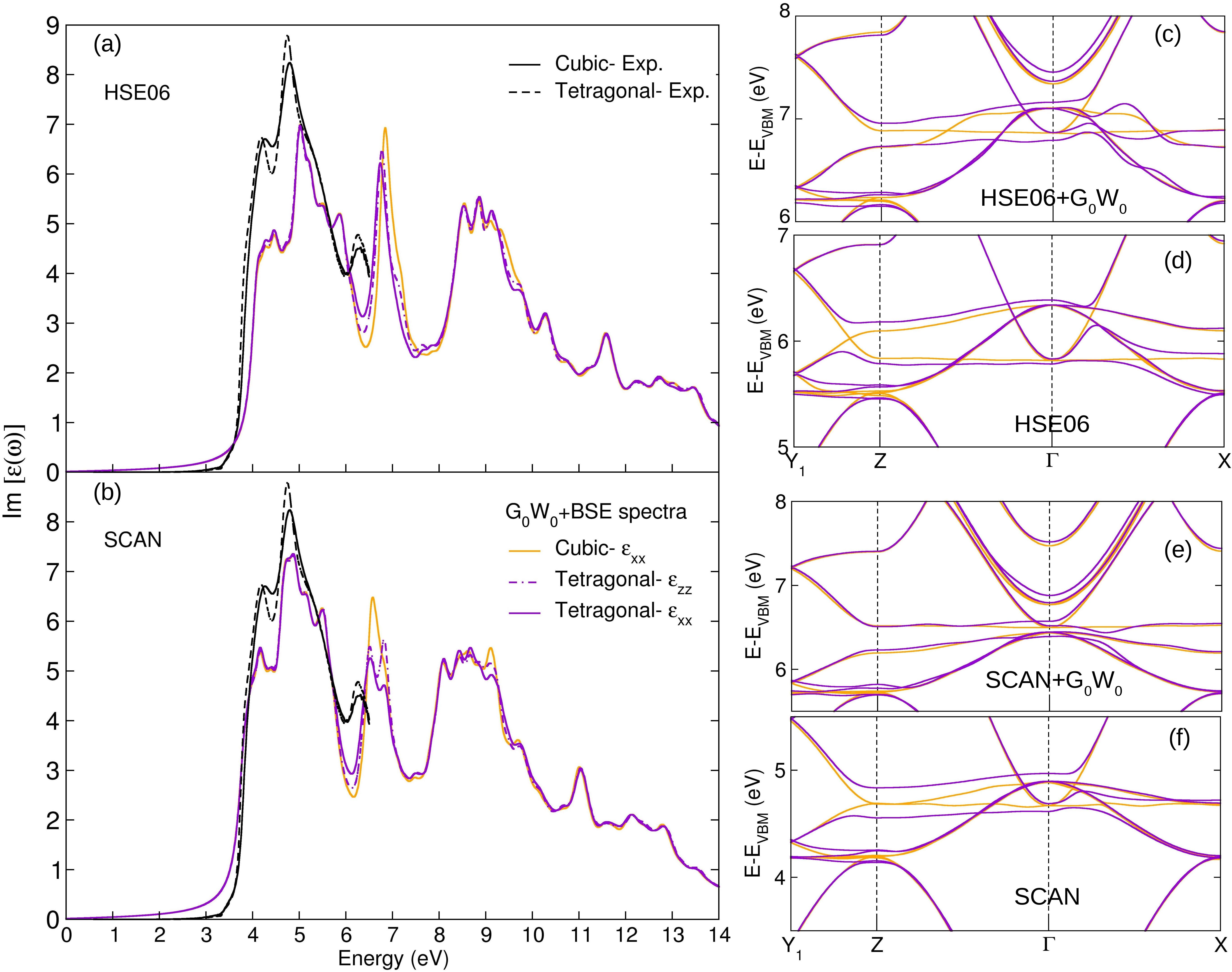}%
\caption{\label{Fig:OptspecAFD}Comparison of the optical spectra after the BSE correction for the cubic and AFD phase of STO for SCAN and HSE06 as starting functionals. The $\varepsilon_{xx}$ (in-plane) and $\varepsilon_{zz}$ (out-of-plane) part of Im[$\epsilon(\omega)$] of the optical spectrum are plotted along with the corresponding band structure in the region around the peak at 6.5 eV to reveal the origin of changes between the tetragonal (purple) and cubic (orange) phases. The experimental spectra from Ref.\citenum{Gogoi} at 4.2 K and 300 K are denoted with black solid and dashed line, respectively.}
\end{figure*}

\subsection{Antiferrodistortive phase}\label{AFD-Opt}

While so far only the optical properties of the cubic phase have been studied  with $G_0W_0$ and BSE \cite{Sponza,Gogoi2}, we explore here the influence of the AFD distortion on the optical spectrum of STO using this approximation. The spectra for the cubic and AFD phases obtained with $G_0W_0$ and BSE for the starting functionals SCAN and HSE06 are plotted in Fig. \ref{Fig:OptspecAFD}a,b. In particular, the  in-plane ($\varepsilon_{xx}$) and out-of-plane ($\varepsilon_{zz}$) components of the AFD phase are compared to $\varepsilon_{xx}$ of the cubic phase. Overall, the spectra for the cubic and tetragonal phases are very similar with an almost indistinguishable onset, due to the very small enhancement in the band gap of 25 meV (SCAN) and 22 meV (HSE06), caused by the small rotational angle of $\sim 2.4\degree$. The main differences are observed around the excitonic peak at $\sim 6.5$~eV, with a slight reduction and splitting of the peak for the SCAN functional and somewhat weaker effect for the HSE06. The observed changes correlate with the modification in the band structure which are most pronounced in this region, as discussed in Section \ref{sec:strel}C. For better comparison we have plotted in Fig. \ref{Fig:OptspecAFD}c-f the HSE06 and SCAN band structures in this region prior and after the $G_0W_0$ calculation. The AFD distortion breaks the symmetry along the $\Gamma-Z$ and $\Gamma-X$ directions. For the latter direction there is a slight shift of the center of mass of weakly dispersive bands towards lower energies which explains the shift of the peak of $\varepsilon_{xx}$ compared to $\varepsilon_{zz}$. 

Temperature-dependent measurements were performed by Gogoi and Schmidt between 4.2 and 300 K in the energy range 0.6-6.5 eV using ellipsometry \cite{Gogoi}. The spectra for 4.2 and 300 K plotted in Fig. \ref{Fig:OptspecAFD}a,b show  a sharpening of the peaks around 4.3 and 6.3 eV at low temperatures, which was attributed to the suppression of vibrational contributions at low temperatures\cite{Gogoi}. Since these are not considered in the  calculation, it is difficult to discern the effect of the structural transition itself from those effects as well as of the possible microstructure formation at low temperatures in experiment.

\section{Conclusion}\label{Sec:Sum}

In summary we have performed a systematic investigation of the role of the exchange-correlation functional on the structural, electronic and optical properties of cubic and tetragonal \sto. SCAN and HSE06 give the best agreement to experiment w.r.t. the structural properties, i.e. lattice constant and rotational angle of the TiO$_6$ octahedra in the AFD phase. SCAN also renders an improved description of the band gap: 2.26 eV vs. 1.81-1.83 eV for PBE96 and PBEsol, respectively. Moreover, SCAN+$U$ with $U=7.45$ eV allows to reach the experimental gap with a moderate enhancement of the lattice parameter (1\%), in contrast to PBE96+$U$ which overestimates the lattice constant by 2\%, whereas the band gap is 0.6 eV lower than the experimental for the same $U$ value.

Concerning the optical properties, the effect of the exchange-correlation functional is gradually reduced by including many body effects. While the difference in the onset of the spectrum amounts to 1.5 eV between PBE96 and HSE06 in the independent particle picture, it is reduced to only 0.3 eV after $G_0W_0$+BSE. $G_0W_0$ is found to overcorrect the band gap by 0.39 eV (PBE96), 0.48 eV (SCAN) and 0.82 eV (HSE06), compared to the experimental value. A good agreement between the theoretical and experimental spectra is obtained only after the solution of the Bethe-Salpeter equation, revealing the crucial role of excitonic effects for STO, in agreement with previous studies based on an LDA functional\cite{Sponza}. Overall the three functionals show good description of the first peak with a slightly higher onset for HSE06. A pronounced excitonic peak appears at $6.4-7.0$ eV depending on the starting functional which can be associated with the  6.3 eV peak in the experimental spectrum. The position of third peak between 8 and 11 eV is best reproduced by HSE06, pointing towards a superior description of the empty Sr $4d$ states with this hybrid functional. Analysis of the oscillator strengths and the electron-hole coupling coefficients plotted in reciprocal space reveal the origin of the main contributions of the peaks in the optical spectrum. Moreover, a model BSE approach is found to give a good description of the main features of the spectrum at a lower computational cost. In general, the mBSE approach represents an interesting alternative in cases where the single particle energies (possibly in combination with the scissors operator) give a good description of the quasiparticle energies. 

Finally, the spectra of the AFD and cubic phases are compared. While the experimental spectrum of Gogoi and Schmidt\cite{Gogoi} shows a sharpening of peaks when the temperature is reduced from 300 to 4.2 K due to the suppression of vibrational contributions, the theoretical spectra reveal the most pronounced effect around the 6.5 eV excitonic peak. This is associated with the changes in band structure due to symmetry breaking along the $\Gamma-X$ and $\Gamma-Z$ directions. Although the observed effects on the optical spectrum are small due to the weak distortion, the methodology and analysis employed here allows deeper insight into the electronic structure of SrTiO$_3$ and is a starting point to address  more complex phenomena such as e.g. polaronic effects in future studies.

\begin{acknowledgments}
We acknowledge useful discussions with G. Kresse, L. Reining, J. Rehr and M. Fechner and funding by the Deutsche Forschungsgemeinschaft (DFG, German Research Foundation) within collaborative research center CRC1242 (project number 278162697, subproject C02) and computational time at the Center for Computational Sciences and Simulation of the University of Duisburg-Essen on the supercomputer magnitUDE (DFG grants INST 20876/209-1 FUGG, INST 20876/243-1 FUGG).
\end{acknowledgments}

\end{document}